\documentclass{article}

\usepackage{PRIMEarxiv}

\usepackage[T1]{fontenc}    
\usepackage{hyperref}       
\usepackage{url}            
\usepackage{booktabs}       
\usepackage{amsfonts}       
\usepackage{nicefrac}       
\usepackage{microtype}      
\usepackage{lipsum}
\usepackage{fancyhdr}       
\usepackage{graphicx}       

\newcommand{\system}{Mojito}

\title{An AI-native Runtime for Multi-Wearable Environments}

\author{
  Chulhong Min, Utku G\"{u}nay Acer, SiYoung Jang, Sangwon Choi, \\
  \textbf{Diana A. Vasile, Taesik Gong, Juheon Yi, Fahim Kawsar} \\
  Nokia Bell Labs \\
}

\begin{document}
\maketitle

\begin{abstract}The miniaturization of AI accelerators is paving the way for next-generation wearable applications within wearable technologies. We introduce \system{}, an AI-native runtime with advanced MLOps designed to facilitate the development and deployment of these applications on wearable devices. It emphasizes the necessity of dynamic orchestration of distributed resources equipped with ultra-low-power AI accelerators to overcome challenges associated with unpredictable runtime environments. Through its innovative approaches, \system{} demonstrates how future wearable technologies can evolve to be more autonomous.
\end{abstract}


\section{Introduction}

\begin{figure*}
\centerline{\includegraphics[width=36pc]{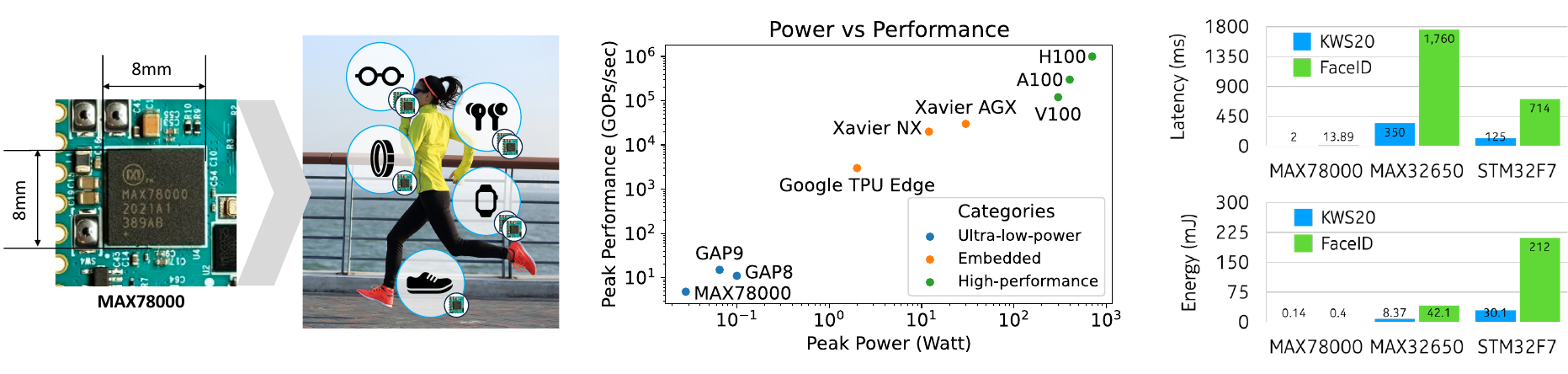}}
\caption{(a) MAX78000 and future wearable computing (left), (b) Comparison of AI accelerators~\cite{accelerators} (middle), (c) Latency and energy comparison between AI accelerator (MAX78000) and microcontrollers (MAX32650 and STM32F7) (right).}\vspace*{-5pt}
\label{fig:max78000}
\end{figure*}

The advent of ultra-low-power AI accelerators, such as Analog MAXIM78000, GreenWaves GAP9, and Google Coral Micro, marks a significant milestone in the evolution of wearable technology. These tiny, yet powerful devices are not just shrinking the physical boundaries of computing but are also redefining the scope of artificial intelligence (AI) on the body. With their tiny form factor (e.g., MAX78000: 8mm$\times$8mm), these accelerators can be seamlessly integrated into wearable devices, transforming our bodies into a network of AI-powered processing units with various sensors, as shown in Figure~\ref{fig:max78000} (a). This integration enables \emph{next-generation wearable applications} to emerge, where distributed wearable devices with ultra-low-power AI accelerators work together to continuously monitor different aspects of human contexts and proactively deliver situational services tailored to the user's immediate needs.

This paradigm shift enables dual-layered collaboration on the user's body. First, it allows wearable applications to flexibly combine different devices for sensing, processing, and delivering information depending on user contexts. For instance, consider an application that employs a smart earbud microphone to help an elderly person with hearing impairments by monitoring surrounding events (e.g., people calling, cars approaching). Then, it can alert the user through a haptic interface on a smart ring worn on the right hand for events occurring on the right side, or through a smartwatch on the left hand for events on that side. A smartwatch’s microphone can be fused for more accurate detection~\cite{cocoon} or dynamically substituted when an earbud is not available. Similarly, heart-related vital signs can be seamlessly monitored by dynamically switching available devices. Second, this ecosystem presents an exciting opportunity to overcome the limitations of the model capacity often faced by single AI accelerators by enabling dynamic collaboration among heterogeneous AI accelerators. Although TinyML techniques, such as model quantization and pruning, enable AI models to fit within a single microcontroller unit (MCU), this compression often leads to significant reductions in accuracy. For instance, compressing the MobileNet model to fit on a single MAX78000 device decreases its accuracy to below 10\%. 

In this context, the role of Machine Learning Operations (MLOps) becomes paramount. MLOps encompasses a broad set of practices aimed at streamlining the lifecycle of machine learning models. The benefit of MLOps includes seamless maintenance and instant deployment of the machine learning models, which are the core workload of these novel wearable applications, ensuring they remain effective and relevant to the user's evolving needs. Despite these advantages, a runtime for wearable devices to facilitate the collaboration of ultra-low-power AI accelerators through MLOps is yet to be developed. We envision \emph{\system{}}, a cutting-edge AI-native runtime for tiny wearable devices equipped with AI accelerators, aimed at enhancing the deployment and execution of next-generation wearable applications in dynamic, resource-constrained environments.

The objective of this article is to examine the convergence of MLOps with next-generation wearable applications. We aim to explore the enabling technology trends and address the research challenges essential for their realization. Furthermore, we will introduce \system{}, an innovative AI-native runtime designed to streamline the development and deployment of these applications. 

\section{TECHNOLOGY TRENDS}

The evolution of next-generation wearable applications is largely driven by three pivotal technology trends: the miniaturization of AI accelerators, the exponential growth of AI models for wearables, and a significant shift in the software paradigm towards AI-centric logic. Each of these trends contributes to the seamless integration and operation of AI within our daily lives, particularly through wearable technology.

\subsection{Miniaturizing AI Accelerators}

At the forefront of this technological wave are ultra-low-power AI accelerators. As shown in Figure~\ref{fig:max78000} (b), these devices mark a significant departure from traditional cloud-scale and edge-scale accelerators, both in terms of size and power efficiency~\cite{ultra-low-power}. Unlike their predecessors, which require substantial power and physical space, these tiny AI accelerators can be embedded directly into small wearable devices.

We introduce the operation principle of these ultra-low-power AI accelerators by taking Analog MAXIM78000~\cite{max78000} as an example. MAX78000 features a dual-core MCU (Arm Cortex-M4 and RISC-V) and a convolutional neural network (CNN) accelerator. The CNN accelerator within the MAX78000 contains 64 convolutional processors, 442 KB of weight storage memory, 2 KB of bias memory, and 512 KB of data memory. These convolutional processors are specially designed for executing convolutional operations at ultra-low-power in parallel; each processor features a pooling engine, input cache, and convolution engine capable of processing up to a 3 by 3 kernel simultaneously. 

A recent benchmark study~\cite{max78000benchmark} demonstrates the MAX78000's superior performance in terms of latency and energy consumption. As shown in Figure~\ref{fig:max78000} (c), the MAX78000 significantly outperforms a conventional microcontroller, MAX32650 with Arm Cortex-M4 and a high-performance controller, STM32F7 with Arm Cortex-M7 in two AI tasks, keyword spotting (KWS) and face detection (FaceID). Latency for KWS is reduced to 2.0 ms compared to 350 ms and 123 ms for MAX32650 and STM32F7, respectively. Energy efficiency is similarly impressive, with MAX78000 consuming only 0.40 mJ for FaceID, in contrast to 42.1 mJ and 464 mJ consumed by MAX32650 and STM32F7.

\subsection{Exponential Growth of AI models for Wearables}

The exponential growth of AI models tailored for wearable devices is driven by the availability of various sensors, not limited to conventional sensors such as camera, microphone and motion sensors, but extending to a wide range of health-related sensors such as Photoplethysmography (PPG), Electrocardiography (ECG), Electroencephalography (EEG), and Electromyography (EMG), and environmental sensors such as temperature, humidity, and air quality sensors. This diversity enables a variety of multi-modal, multi-sensory AI models to emerge. In addition to well-established AI models such as object detection, keyword spotting and physical activity detection, they can provide comprehensive insights into a user's health and surrounding events. For example, a model could analyze data from PPG and ECG sensors to offer real-time heart health monitoring, while another could use EEG and EMG data to detect stress levels or neurological conditions. Moreover, the synergy between different sensor modalities can significantly enrich AI models, allowing them to effectively capture and fuse complementary information from different modalities~\cite{multi-modal}.

\subsection{Changing Paradigm of Wearable Applications}

The third trend reshaping the landscape is the changing paradigm of wearable applications. Historically, wearable applications have relied on heuristic and straightforward logic for their functionality, focusing on basic task execution and data presentation. However, with advancements in AI accelerator miniaturization and the exponential growth of AI models, the next-generation of wearable applications will adopt AI as its core logic. This shift toward \emph{data-driven AI} allows wearable devices to not just perform predefined, simple tasks but also adapt and respond intelligently to the user's specific contexts and needs. This transition will enable wearables to deliver more personalized, context-aware experiences by harnessing the rich data collected through various sensors.

\section{CHALLENGES FOR WEARABLE COLLABORATION}

The emerging technology trends significantly empower wearable applications through dual-layered collaboration of AI accelerator-equipped wearables, but they also introduce complex challenges that need to be addressed at the system level. While extensive research efforts have been made in the domain of MLOps to streamline the lifecycle of machine learning models, they primarily focus on \textit{model}-centric view on a \textit{single} device and often overlook holistic execution support from the \emph{application}'s perspective. Moreover, dual-layered collaboration in wearable-rich environments brings new and exciting research challenges, such as tight time synchronization across distributed sensors and navigating heterogeneous and dynamic runtime environments. Addressing these issues is crucial for realizing the full potential of next-generation wearable applications.

\begin{figure}
\centerline{\includegraphics[width=15pc]{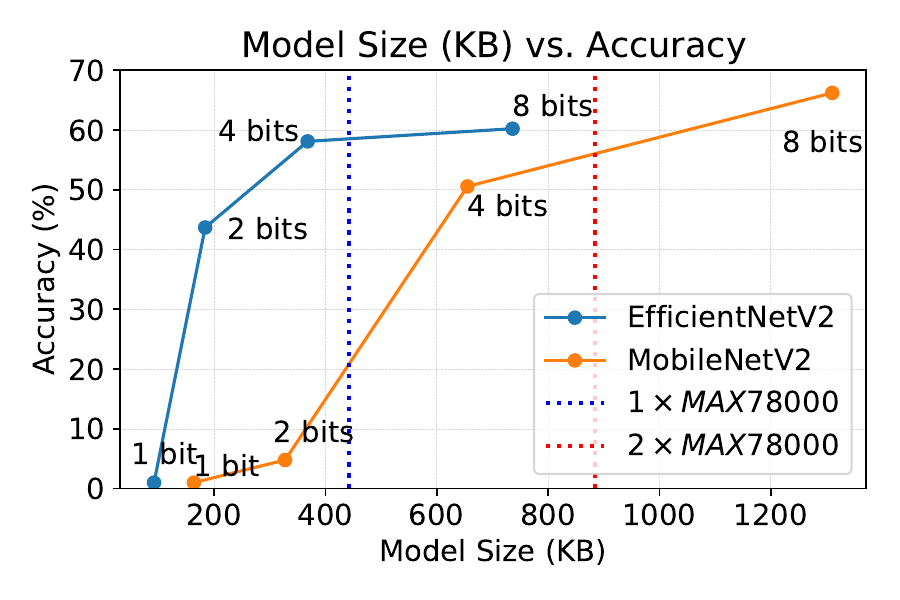}}
\caption{Accuracy variation according to quantization bit (1, 2, 4, 8) size of EfficientNetV2 and MobileNetV2.}\vspace*{-5pt}
\label{fig:challenges}
\end{figure}

\subsection{Bottleneck of Ultra-low-power AI Accelerators: Memory}

The adoption of TinyML techniques~\cite{tinyml}, aimed at facilitating AI miniaturization, encounters obstacles primarily due to memory constraints of ultra-low-power AI accelerators. Techniques like model quantization and pruning, designed to fit AI models into the extremely limited memory of MCUs, can severely impact the performance of more complex models. For example, quantizing MobileNet to fit into MAX78000, which supports up to 442KB of AI model weight memory, drastically reduces its accuracy to under 10\%, as shown in Figure~\ref{fig:challenges}. However, we can achieve higher accuracy by utilizing additional memory capacity through the collaboration of multiple 78K devices. If three devices are available, they can collaboratively afford 8 bit-quantized MobileNet, showing 65\% of the accuracy. This compromise between model complexity and capacity highlights a critical bottleneck in leveraging the full potential of AI in wearable devices, and motivates the need for the collaboration of ultra-low-power AI accelerators.

\subsection{Tight Time Synchronization}

A critical challenge for wearable collaboration, particularly those utilizing multi-modal AI models, is ensuring tight time synchronization. Even tens of milliseconds of synchronization error can significantly detract from the performance of AI models. This issue becomes more challenging, especially in wearables that are distributed and wirelessly connected, where achieving precise time synchronization presents additional hurdles. Some small wearable devices lack a real-time clock, inherently limiting their ability to maintain accurate time. Also, energy-saving measures often result in intermittent network interface availability, further complicating synchronization efforts. 

\subsection{Heterogeneous and Dynamic Runtime Environment}

The landscape of wearable devices and AI accelerators is diverse and dynamic, presenting significant challenges for developers in leveraging device collaboration. For instance, smart earbuds only become available for use when worn, and smartwatches may become unavailable due to battery depletion. This variability makes predicting the runtime environment difficult, limiting developers' ability to integrate and manage the capabilities of available devices effectively. Such unpredictability hinders the seamless composition of device functionalities, which is essential for delivering responsive and user-centric services.


\section{\system{}: An Overview}

Current wearable technology operates within a critical limitation: the absence of a runtime system designed to support the collaborative functionality of wearable devices while addressing the challenges outlined previously. Most minimal wearables today rely solely on firmware, meaning their functionalities are confined to predefined tasks, with updates possible only through firmware upgrade. This approach significantly limits the wearables' adaptability and functionality, as they cannot dynamically respond to the runtime environment or extend their capabilities beyond the singular device. This lack of an advanced runtime environment restricts the emergence of next-generation wearable applications.

We envision \system{} as an AI-native runtime that seamlessly integrates wearable technology with AI, aiming to facilitate the development and deployment of next-generation wearable applications. The fundamental concept behind \system{} is the separation of device collaboration from AI models. This means that \system{} accepts AI models in their unmodified state (optimized during the training phase) and ensures their optimal execution by dynamically composing distributed AI accelerators on wearables. Here, we introduce the proposed innovative features to realize \system{}.

\textbf{Virtualizing Dynamic Resources}: At its core, \system{} simplifies the complex job of integrating and managing a collection of devices by virtualizing dynamic resources such as sensors, AI processors, and interfaces. It creates a unified abstraction layer that allows wearable applications to access and utilize a collective pool of distributed resources as if they were a single, powerful device. This virtualization allows developers to focus on application logic without having trouble addressing the complexities of a variable deployment environment.

\textbf{Orchestration of Heterogeneous AI Accelerators}: A key strength of \system{} is its capability to orchestrate heterogeneous AI accelerators efficiently. It intelligently breaks down AI models into smaller subtasks, assigning them to the most suitable AI accelerators based on current demands and resource availability. Through dynamic adjustment of AI accelerator configurations, \system{} achieves the system-wide performance, e.g., maximizing the model execution throughput or minimizing the total energy consumption.






\section{VIRTUAL COMPUTING SPACE}

\begin{figure*}
\centerline{\includegraphics[width=35pc]{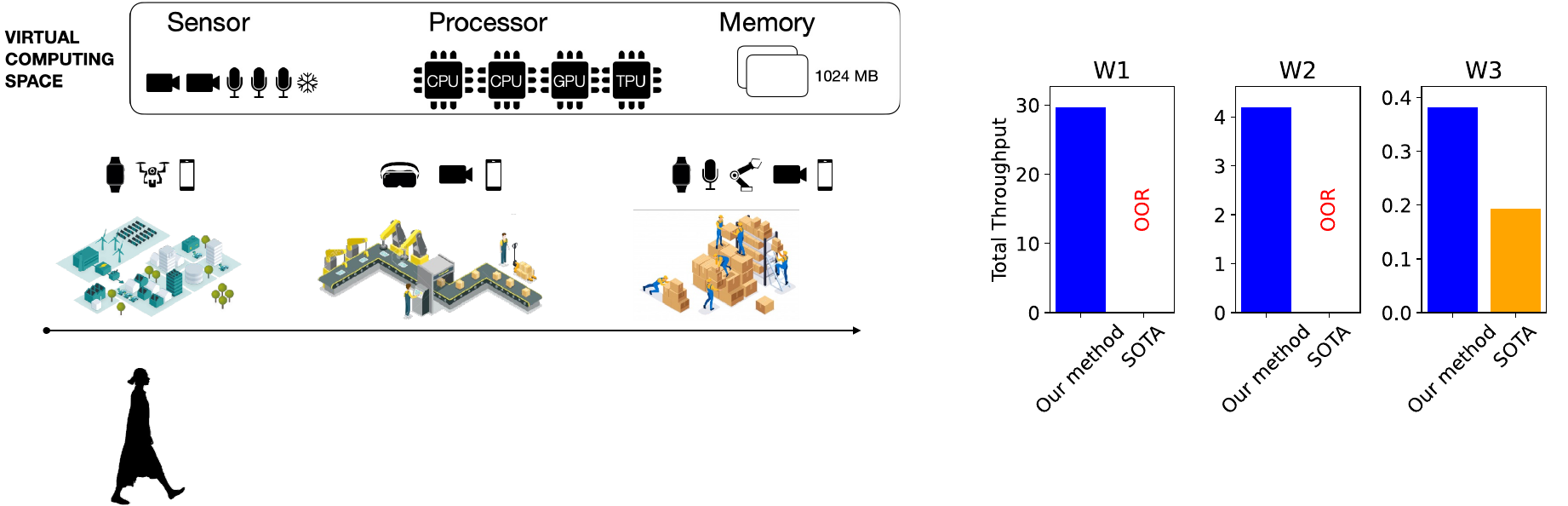}}
\caption{(a) Illustrative example of virtual computing space (left), (b) Throughput comparison with state-of-the-art method; W1: ConvNet, ResSimpleNet, UNet, W2: KeywordSpotting, SimpleNet, WideNet, W3: EfficientNetV2 (right).}\vspace*{-5pt}
\label{fig:system}
\end{figure*}

We propose a \emph{virtual computing space}, a conceptual layer offering a unified and virtual perspective on the computational resources distributed across AI accelerator-equipped wearables. As shown in Figure~\ref{fig:system} (a), as a user navigates throughout a day, the collection of available computing resources dynamically changes--encompassing not only the wearables on the body but also nearby IoT devices. It provides application developers with a unified view as if distributed heterogeneous resources were a single device by masking the devices' diverse and dynamic nature. Within the virtual computing space, application logic employing virtual resources is dynamically translated to physical ones by an intelligent orchestration layer, tailored to AI processing demands and device capabilities. This approach enhances resource efficiency, ensures operational consistency despite fluctuating device availability, and facilitates scalability.

\subsection{Application Programming Interface}

To realize virtual computing space, we introduce a novel programming abstraction for distributed wearable systems. Different from traditional MLOps platforms, which often focus mostly on the support to \textit{model} execution, \system{} allows for the specification of complete pipelines—from data collection to processing the outcomes of model computations. To handle the unpredictable and varied nature of wearable ecosystems, the key proposed idea is to separate pipeline logic from specific hardware requirements. For model operations, applications define what needs to be run and our orchestration framework then intelligently allocates these tasks to the most suitable AI accelerators available. For sensing tasks, \system{} intelligently selects the most appropriate sensing device based on the application's specified needs, such as sensor type or location. The criteria for output destinations are similarly flexible, allowing specifications based on interface type or physical location. 

We introduce two primary functions: \textit{\textbf{register()}} and \textit{\textbf{unregister()}} to register and unregister on-body proactive AI applications, respectively. An application can be described by sensing needs, the model to be executed, the post-processing function, and output requirements. For example, an application that monitors heart-related vital signs and alerts anomaly status via earbud can be defined as (PPG, HeartAnalysis, \textit{anomalyDetection()}, earbud). Similarly, an application that detects nearby person's calling event can be defined as (microphone, KeywordSpotting, \textit{vibrate()}, haptic). 

\section{AI ACCELERATOR ORCHESTRATION}

We develop a runtime orchestration module designed to seamlessly translate virtual tasks from application logic into physical hardware resources~\cite{synergy}. Here, we focus on the model execution and its distribution across heterogeneous AI accelerators. In wearable environments, a traditional way to run AI models on resource-constrained MCUs is to separately compress the model to fit the device where the model runs. However, as shown in Figure~\ref{fig:challenges}, this often leads to the model becoming unusable due to the highly limited memory capacity of MCU devices. Also, more importantly, such a model manipulation approach (i.e., adjusting models to fit available devices) is not practical in environments with heterogeneous devices since the compression operation needs to be done repeatedly for each device.

To address this issue, we propose a novel concept, \emph{AI accelerator manipulation} for model execution support. This method leverages available AI accelerators through collaborative inference, allowing the input model to be executed without any modifications at runtime. It optimally distributes the tasks required for the model execution across a network of heterogeneous AI accelerators. The key enabler of this strategy is three-fold: First, we introduce a systematic method for generating execution plan candidates from application logic with dynamic resource availability. This approach allows \system{} to have more flexible and efficient task mapping. Second, we extend beyond the conventional focus on partitioning the AI model into tasks. We identify that data transfer overhead between sensing and information processing devices is critical in wearable computing and devise the source-target-aware orchestration approach; e.g., placing the early layers of the model closer to the corresponding sensor while avoiding data congestion with other models. Last, we propose an online latency prediction technique, specially designed to accommodate the unique memory operations and processing architecture of ultra-low-power AI accelerators.

Our evaluation, conducted on Analog MAX78000 and MAX78002 platforms, demonstrates its superiority over existing model partitioning techniques. \system{} shows remarkable performance improvements, i.e., achieving an average of 8.0$\times$ model throughput compared to state-of-the-art methods~\cite{neurosurgeon}; Figure~\ref{fig:system} (b) shows the detailed results for three scenarios with four Analog MAX78000 devices; OOR represents out-of-resource failures due to resource conflict among models. Furthermore, the results reveal the adaptability of \system{} to changes in the runtime environment, including device availability and resource heterogeneity.
\section{Exploring Untapped Opportunities}

We have concentrated on research challenges and solutions associated with runtime device collaboration for next-generation wearable applications. However, through the exploration and implementation of \system{}, we have identified additional issues that call for attention but have yet to be thoroughly examined in the research community. Among these, we introduce two issues that can be interesting opportunities for the future \system{}.

\subsection{On-the-fly Device-to-Device Authentication}

The dynamic collaboration among on-body devices introduces challenges in secure and efficient authentication on-the-fly~\cite{proteus}. Many wearables, due to their miniaturization, lack the hardware for biometric identification or touchscreens for passcode entry, often relying on one-time Bluetooth associations. This limitation is problematic for next-generation wearable applications that depend on the dynamic collaboration of available devices, as the lack of on-the-fly device-to-device authentication could compromise user privacy and model performance. For example, smart earbuds cannot detect shared usage contexts, leading to potential privacy breaches and inaccuracies in health data collection. Additionally, with AI applications becoming new stakeholders in device collaboration, there's a risk that malicious models or devices could expose privacy-sensitive health data. This limitation emphasizes the need for developing more adaptable and secure authentication strategies that can accommodate the unique constraints and usage patterns of wearable devices.

\begin{figure}
\centerline{\includegraphics[width=15pc]{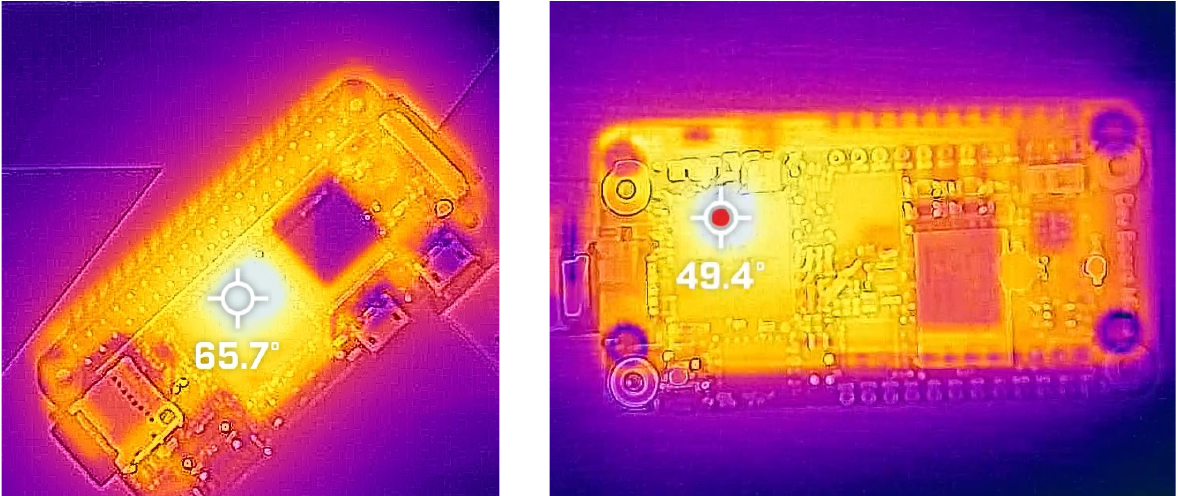}}
\caption{Temperature measurement of Raspberry Pi Zero (left) and Google Coral Micro with AI processing (right).}\vspace*{-5pt}
\label{fig:thermal}
\end{figure}

\subsection{Thermal Comfort on On-body AI}

In on-body AI, thermal comfort becomes a paramount requirement due to the continuous nature of AI execution. The close proximity of devices to the user's skin elevates the importance of maintaining thermal safety to prevent discomfort or harm. The challenge of effective thermal management becomes evident, especially as AI workloads cause device temperatures to rise, potentially leading to performance degradation through thermal throttling. For instance, Figure~\ref{fig:thermal} shows that small form-factor single-board computers like Raspberry Pi Zero and Google Coral can quickly reach temperatures up to around 50°C-60°C. This poses a significant concern as continuous exposure to temperatures over 42°C-45°C can lead to thermal-regulatory disorders and serious health risks such as thermal pain, skin burns, and skin aging~\cite{thermal}. This scenario highlights the critical need for advanced thermal regulation strategies to ensure user safety and device performance in wearable technologies.

\section{Conclusion}

We introduced exciting research challenges that MLOps need to address for next-generation wearable applications. \system{} highlighted the importance of dynamic and holistic orchestration of wearable devices. Moving forward, we will shift our focus to exploring other issues in the context of automated model execution, e.g., efficient memory management for collaborative inference, system-driven autonomous model personalization, and expansion of device collaboration to multi-user environments. We will integrate these solutions into \system{}. We aim to deepen our understanding of these complex systems, moving towards a future where on-body AI enhances our lives by providing us with support in every moment of our lives.

\bibliographystyle{unsrt}  
\bibliography{references}

\end{document}